%% file: ms.tex
\documentclass[reprint, byrevtex, aip, jap, amsmath, amssymb, showkeys,
  superscriptaddress, citeautoscript, floatfix]{revtex4-2}
\pdfoutput=1

\usepackage[utf8x]{inputenc}
\usepackage[OT1]{fontenc}
\usepackage[USenglish]{babel}
\usepackage[table]{xcolor}      
\usepackage[pdftex]{graphicx}   
\usepackage[inline]{enumitem}
\usepackage{dcolumn}            
\usepackage{multirow, tabularx} 
\usepackage{bm}                 
\usepackage{makecell}           
\usepackage{natbib}
\usepackage{xspace}
\usepackage{siunitx}            
\usepackage{hyperref}

\hypersetup{
    pdffitwindow=false,     
    pdfstartview={FitH},    
    pdftitle={Robustness of Parameter Estimation Procedures for Bulk-Heterojunction Organic Solar Cells},    
    pdfauthor={A. Prel, A. Rezgui, A.-S. Cordan and Y. Leroy},     
    pdfkeywords={Organic Solar Cells, Bulk Hetero-Junction, Bayesian, Parameter estimation}, 
    pdfnewwindow=true,      
    colorlinks=true,       
    linkcolor=blue,          
    citecolor=blue,        
    filecolor=black,         
    urlcolor=blue        
}

\DeclareSIUnit{\muUnit}{\meter\squared\per\volt\per\second}  
\DeclareSIUnit{\GUnit}{1\per(\centi\meter\cubed\second)}     
\DeclareSIUnit{\kUnit}{\meter\cubed\per\second}              
\DeclareSIUnit{\points}{\mathrm{points}}

\def\phi{\varphi}

\def\vec{\bm}

\def\etal{\emph{et~al{.}}}
\def\None{}
\def\ee{\,{=}\,}
\def\CV{\mbox{C-V}\xspace}
\def\IV{\mbox{I-V}\xspace}
\def\Cf{\mbox{C-freq}\xspace}
\def\GT{GT\xspace} 

\newcommand{\tref}{\mathrm{ref}}
\newcommand{\est}{\mathrm{est}}
\newcommand{\Yref}{Y_\tref}
\newcommand{\given}{\,{\mid}\,}  

\newcommand{\epsr}{\varepsilon_r}
\newcommand{\Cgeom}{C_{\mathrm{geom}}}
\newcommand{\phicat}{\phi_{\mathrm{cat}}}
\newcommand{\phian}{\phi_{\mathrm{an}}}
\newcommand{\Vbi}{V_{\mathrm{bi}}}
\newcommand{\Vgap}{V_{\mathrm{gap}}}
\newcommand{\Geff}{G_{\mathrm{eff}}}
\newcommand{\Gopt}{G_{\mathrm{opt}}}
\newcommand{\krec}{k_{\mathrm{rec}}}

\newcommand{\JSC}{J_\mathrm{SC}}
\newcommand{\VOC}{V_\mathrm{OC}}

\newcommand{\Vpeak}{V_{\mathrm{peak}}}  

\def\ICube{%
  \affiliation{
    ICube Laboratory, CNRS\,/\,Université de Strasbourg, Télécom Physique Strasbourg,
    Illkirch, France.
  }%
}

\def\ESYCOM{%
  \affiliation{
    ESYCOM, CNRS\,/\,Université Gustave Eiffel, ESIEE Paris, Marne-la-Vallée, France.
  }%
}

\begin{document}

  \title{Robustness of Parameter Estimation Procedures for Bulk-Heterojunction Organic Solar Cells}

  \author{Alexis \surname{Prel}}
  \ICube
  
  \author{Abir \surname{Rezgui}}
  \ESYCOM

  \author{Anne-Sophie \surname{Cordan}}
  \ICube

  \author{Yann \surname{Leroy}}
  \email[Correspondence to: ]{yann.leroy@unistra.fr}
  \ICube

  \date{\today}

  \begin{abstract}

    Parameter estimation procedures provide valuable guidance in the
    understanding and improvement of organic solar cells and other devices.  They
    often rely on one-dimensional models, but in the case of bulk-heterojunction
    (BHJ) designs, it is not straightforward that these models' parameters have a
    consistent physical interpretation.  Indeed, contrarily to two- or
    three-dimensional models, the BHJ morphology is not explicitly described in
    one-dimensional models and must be implicitly expressed through effective
    parameters.
    In order to inform experimental decisions, a helpful parameter estimation
    method must establish that one can correctly interpret the provided parameters.
    However, only a few works have been undertaken to reach that objective in
    the context of
    BHJ organic solar cells.  In this work, a realistic two-dimensional model of
    BHJ solar cells is used to investigate the behavior of state-of-the-art
    parameter estimation procedures in situations that emulate experimental
    conditions.
    We demonstrate that fitting solely current-voltage characteristics by an
    effective medium one-dimensional model can yield nonsensical results, which may
    lead to counter-productive decisions about future design choices.  In agreement
    with previously published literature, we explicitly demonstrate that fitting
    several characterization results together can drastically improve the robustness
    of the parameter estimation.  Based on a detailed analysis of parameter
    estimation results, a set of recommendations is formulated to avoid the most
    problematic pitfalls and increase awareness about the limitations that cannot
    be circumvented.

  \end{abstract}


\maketitle

\section{Introduction}

\begin{figure}[t]
  \centering%
  \includegraphics[width=\linewidth]{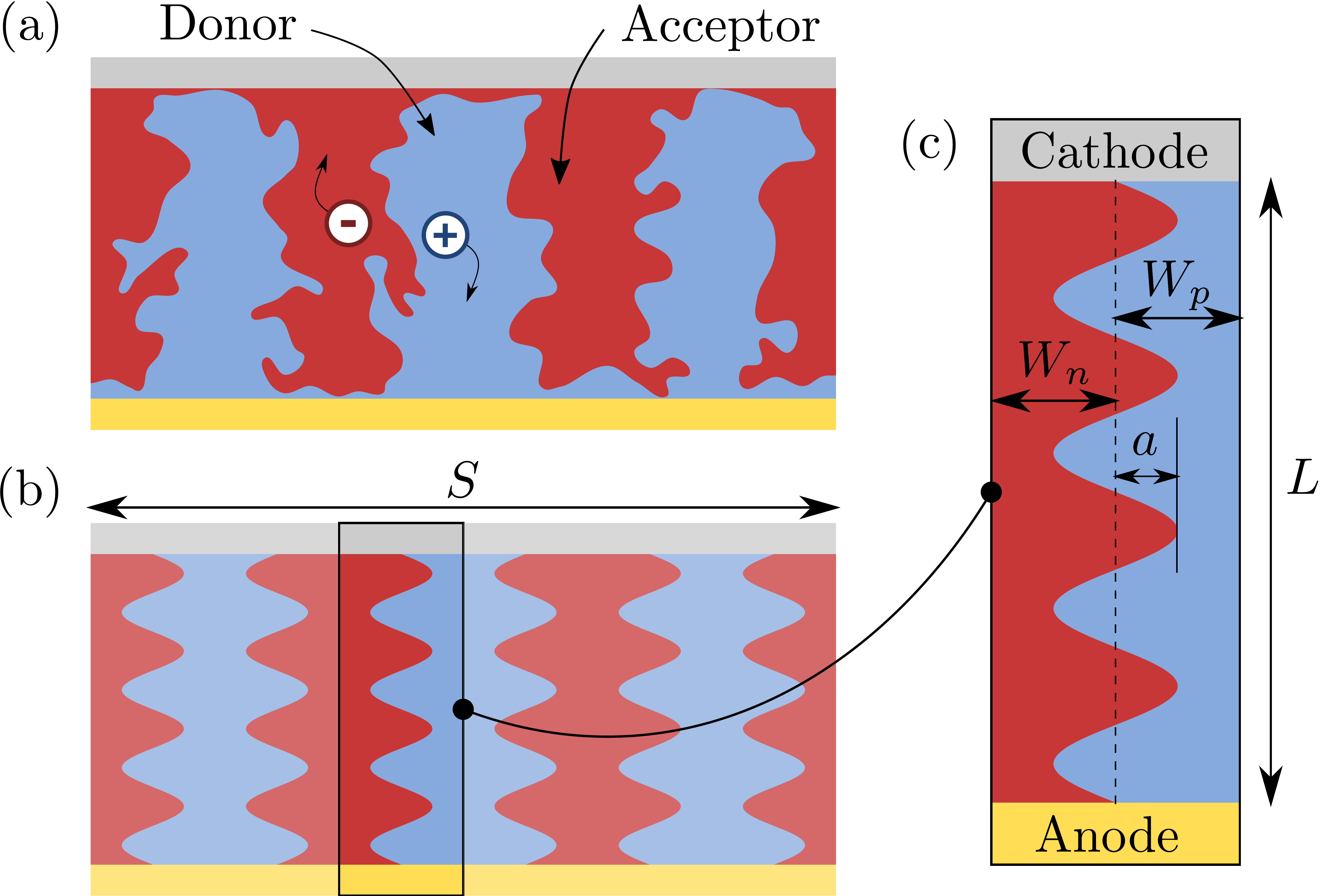}%
  \caption{\label{fig:geometries}%
    a)~Complex percolation pathways observed in actual BHJ architecture;
    b)~Model percolation pathways.  Their tortuosity are controlled by a
    sinusoidal interface.  The active area $S$ is reduced to an elementary
    structure, repeated using mirroring boundary conditions;
    c)~Parameters of the elementary structure considered for the simulations:
    active layer thickness $L$, half-widths of the material pathways $W_{n,p}$, 
    and amplitude $a$ of the sinus interface made of $N_\mathrm{per}$ periods.
  }%
\end{figure}

The world's renewable energy consumption is projected to keep on rising during
the upcoming decades~\cite{EIA_international_2019,kober_global_2020}.
Among other means of production that could help supply that demand, Organic 
Solar Cells (OSC) are a promising technology, but their power conversion 
efficiency remains lower than competing technologies~\cite{green_solar_2019}.

One of the most widespread strategies to improve OSC performances is
the so-called bulk-heterojunction (BHJ) architecture for the active layer.
In a BHJ, a donor and acceptor materials are finely mixed to form
bicontinuous percolation pathways for free carriers, as depicted in
Fig.~\ref{fig:geometries}a.
Such a complicated morphology is difficult to probe with traditional investigation tools.
Therefore, to further improve BHJ devices, experimental exploration would
benefit from reliable modeling insight that can operate on a reasonable
computational budget and explain the inner optoelectronic device processes
that are hard to probe directly.  To this end, parameter estimation is a helpful
diagnostic tool to interpret routine device
measurements~\cite{zonno_extracting_2009}.

However, BHJ geometries are more challenging to model than a
planar stack of layers.
In theory, one should model the entire three-dimensional
(3D) morphology and account for the influence of tortuous conduction pathways on
transport to reproduce experimental observations~\cite{albes_investigation_2016}.
In practice, detailed information at the relevant length-scale (${\simeq}\,
\SI{10}{\nano\meter}$) is seldom available.  As a result, it is often argued
that the active layer can be modeled as a one-dimensional (1D) effective medium
with uniform physical properties, which result from a combination of the
characteristics of the two considered materials and the BHJ morphology
itself~\cite{bartesaghi_charge_2014,hwang_modeling_2008}.
While it may be possible to carefully design 1D models to match the predictions
of a more realistic model~\cite{richardson_derivation_2017}, the physical
properties of the effective medium are typically chosen to match the
current-voltage (\IV) response of a fabricated device~\cite{koster_device_2005,%
koster_bimolecular_2006,albes_investigation_2016,tang_relating_2018}, which can
lead to arbitrary values that are not reliable \cite{set_what_2015,%
albes_investigation_2016,neukom_opto-electronic_2018}.  However, it is highly
desirable for the fit parameters to predict measurements not exploited during
inference, identify performance bottlenecks, or otherwise suggest appropriate
experiments to try next~\cite{set_what_2015}.  This work presents pieces of
evidence, through numerical studies, that challenge the value of a 1D effective
medium model to reach these goals.

As of today, the most realistic models such as master equation 
\cite{li_modeling_2018, van_modeling_2009},
kinetic Monte  Carlo
\cite{albes_investigation_2016,gagorik_monte_2013,wilken_experimentally_2020},
or molecular dynamics approaches
\cite{martinelli_modeling_2009},
are prohibitively expensive for model-intensive applications such as parameter
estimation \cite{raba_organic_2017,neukom_opto-electronic_2018}
or machine-learning
\cite{majeed_using_2020}.
In contrast, fitting is often performed via equivalent circuit analysis
\cite{jordehi_parameter_2016,liao_parameters_2016} using various algorithms
\cite{li_evaluation_2013,chin_cell_2015}, but the parameters derived in this
way are difficult to relate to the internal physics of the active layer.
Here, we focus on drift-diffusion models, as they offer a good compromise
between the granularity of the description and the computation time required for
simulations\cite{groves_simulating_2016}, while keeping parameters with physical
meanings.

With this kind of model, existing fitting approaches provide reasonable
parameter values when applied to experimental measurements.  However, the values
are not guaranteed to be relevant if they describe a 1D effective medium
model~\cite{koster_device_2005,%
koster_bimolecular_2006,albes_investigation_2016,tang_relating_2018}, nor unique
if obtained by local optimization~\cite{raba_organic_2017}.
To test fitting procedure robustness, we propose a reliable and fully reproducible
assessment protocol.  The protocol is based on a synthetic dataset of optoelectronic
characterizations generated by a two-dimensional (2D)
drift-diffusion model~\cite{raba_organic_2014}.
We apply it to two fitting procedures: the direct fitting of \IV characteristic
alone and a more complex method that considers multiple characterizations~%
\cite{neukom_opto-electronic_2018}. To identify possible multiple local optima
and discuss error bars, we work with Bayesian tools as in a previous
contribution~\cite{raba_organic_2017}.

We demonstrate that parameters extracted solely from the \IV curve are not
reliable and we illustrate with a clear example the typical misinterpretation
that may arise.  Then, we show that many of the issues identified with this
approach can be mitigated if one considers more than a single characterization
technique~\cite{neukom_opto-electronic_2018}.
After identifying and discussing some limitations of the procedure given in
Ref.~\onlinecite{neukom_opto-electronic_2018}, we present an improved
procedure that requires fewer optoelectronic characterizations for the fit and
obtains better agreement to the data, even for characterizations not exploited
for inference.
As none of the evaluated procedures retrieve the values of the parameters chosen
for the 2D model, the physical meaning of the effective parameters obtained is
discussed.

\section{Methodology}
\label{sec:methodo}

\subsection{Robustness evaluation protocol}
\label{sec:protocol}

\begin{figure}[b]
  \centering
  \includegraphics[width=1.0\linewidth]{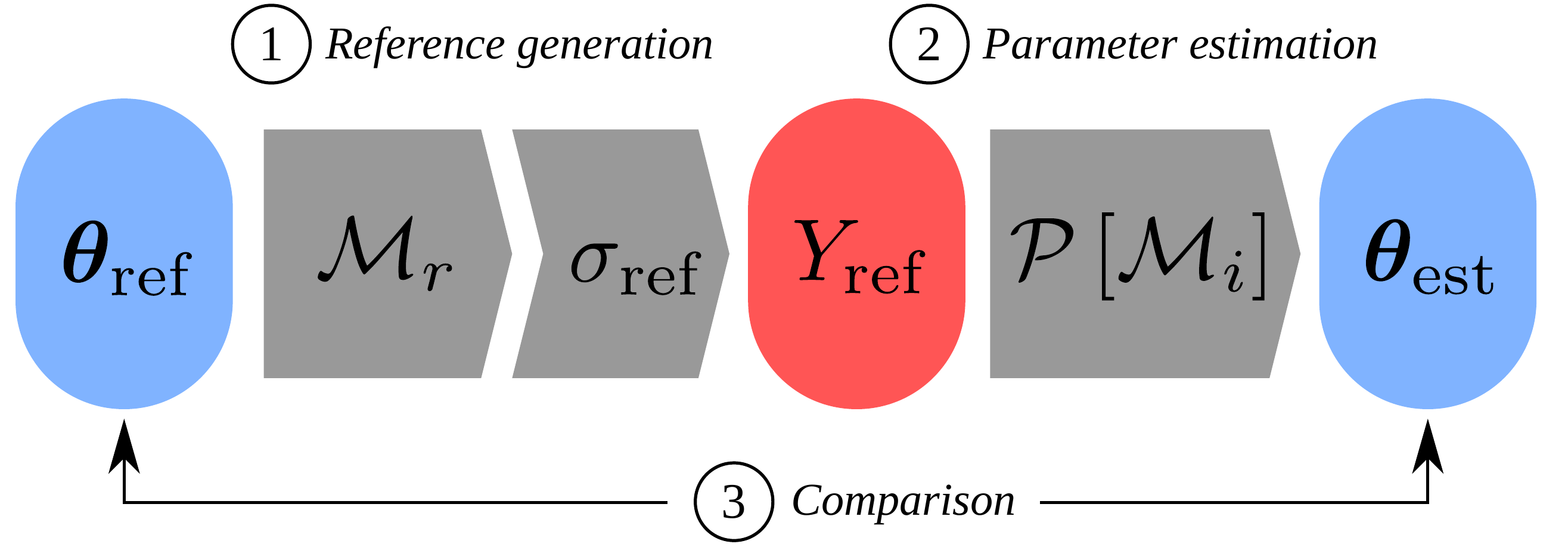}
  \caption{\label{fig:protocol}%
    Evaluation protocol for parameter estimation procedure
    $\mathcal{P}\left[\mathcal{M}_i\right]$.  Inferred parameters $\vec\theta_\est$
    are compared to the ground-truth $\vec\theta_\tref$ taken to generate synthetic
    data $\Yref$ from model $\mathcal{M}_r$ and apparatus noise $\sigma_\tref$.
  }
\end{figure}
Proving that a fitting procedure is trust-worthy is difficult, because an
independent validation of the parameters obtained is typically lacking.
However, it is easy to confirm that this procedure performs adequately in a
simulated context where the ground-truth (GT) is known.

In this work, the robustness of several parameter estimation procedures is
evaluated according to a protocol depicted in Fig.~\ref{fig:protocol}: 
1) ground-truth parameters $\vec\theta_\tref$ are chosen to generate the
reference data $\Yref$, from a model $\mathcal{M}_r$ considered at least as
realistic as the model $\mathcal{M}_i$ used for inference; 
2) a parameter estimation procedure $\mathcal{P}\left[\mathcal{M}_i\right]$ is
applied to $\Yref$; and
3) the inferred parameters $\vec\theta_\est$ are compared to $\vec\theta_\tref$.
This validation is necessary but insufficient because the reference-generation
model $\mathcal{M}_r$ may fail to account for processes that influence real
measurements.
\begin{table*}
  \caption{%
    \label{tab:characterization_list}%
    Characterization data included in the reference datasets.  For this work we
    follow the definitions from Ref.~\onlinecite{neukom_opto-electronic_2018}.
  }%
  \begin{ruledtabular}%
    \begin{tabular}{llllr}%
	    Abbreviation & Measurement & Conditions & Apparatus & Ref. \\[1pt]
      \hline \\[-0.8em]
	    dark \IV & \IV curve & In the dark.
        & Keithley 2420  & \onlinecite{keithley2400} \\[1pt]
	    light \IV & \IV curve & Under 1 sun illumination.
        & Keithley 2420  & \onlinecite{keithley2400} \\[1pt]
	    dark-CELIV & CELIV & In the dark.
        & R{\&}S RTM3004 & \onlinecite{oscilloscope} \\[1pt]
	    photo-CELIV & CELIV & After exposure to 1 sun illumination.
        & R{\&}S RTM3004 & \onlinecite{oscilloscope} \\[1pt]
      TPC & Transient Photo-Current & Switch from 0 to 1 sun illumination at
        short-circuit. & R{\&}S RTM3004 & \onlinecite{oscilloscope} \\[1pt]
      \Cf & Capacitance-frequency & AC perturbation around short-circuit
        conditions. & Agilent 4294A  & \onlinecite{agilent4294a} \\[1pt]
      dark \CV & Capacitance-voltage & AC perturbation around each point of the
        dark \IV curve. & Agilent 4294A  & \onlinecite{agilent4294a} \\[1pt]
      TPV & Transient Photo-Voltage & Switch from 1 to 0 sun illumination at
        open-circuit. & R{\&}S RTM3004 & \onlinecite{oscilloscope}
    \end{tabular}%
  \end{ruledtabular}%
\end{table*}%

The synthetic reference data $\Yref$ is generated using a realistic 2D model,
denoted by $\mathcal M_r$ in Fig.~\ref{fig:protocol}, already reported in
another publication \cite{raba_organic_2014}.  In contrast, the fitting
procedures $\mathcal{P}$ rely on a 1D effective medium model $\mathcal{M}_i$, as
described for instance in
Refs.~\onlinecite{koster_device_2005,neukom_opto-electronic_2018}. To simulate
$Y$ and $\Yref$, both models were implemented in the finite elements software
COMSOL Multiphysics\textsuperscript{\textregistered} \cite{Comsol}. $\Yref$ is
comprised of a set of up to eight synthetic measurements, using six
characterization techniques, summarized in
Table~\ref{tab:characterization_list}.
To account for the measurement noise, a perturbation is added to $\Yref$, drawn
from a normal distribution with zero-mean and a standard-deviation
$\sigma_\tref$ taken from datasheet specifications of each apparatus
\cite{keithley2400,agilent4294a,oscilloscope}.

\subsection{Bayesian inference}
\label{sec:bayesian_inference}

In least-square fitting procedures, the root mean square error (RMSE)
quantifies the disagreement between a reference dataset $\Yref$ and the
prediction $Y(\vec\theta)$ of a fitting model $\mathcal{M}_i$. It is to be
minimized with respect to the model's parameters $\vec\theta$, and is defined as
\begin{equation}
	\label{eq:MSE_expression}
	\mathrm{RMSE}(\vec\theta) = \sqrt{\frac{1}{m}\sum_{j=1}^{m} w_j \left(
	    Y_{j}(\vec\theta) - Y_{\tref, j}
	\right)^2}
\end{equation}
where $\Yref \ee \{ Y_{\tref, j},\: j \in 1, \ldots, m \}$ are $m$ reference
datapoints, $Y(\vec\theta) \ee \{Y_{j},\: j \in 1, \ldots, m \}$ is the
corresponding prediction of $\mathcal{M}_i$, and $ w \ee \{w_j,\: j \in 1,
\ldots, m \}$ are weighting factors.

Least-square fitting can be seen as a special case of the Bayesian approach
to parameter estimation.  Instead of the RMSE, the Bayesian picture considers
the posterior probability density $p(\vec\theta \given \Yref)$, the probability
density of $\vec\theta_\tref$ being equal to $\vec\theta$, posterior to the
experimental observation $\Yref$.  It is given by
\cite{mackay_information_nodate}
\begin{equation}
	\label{eq:Bayes_theorem}
	p(\vec\theta \given \Yref) \propto p(\Yref \given \vec\theta)
	                           \times p(\vec\theta)
\end{equation}
where $p(\Yref \given \vec\theta)$ is the likelihood of observing $\Yref$, if
the hypothesis $\vec\theta=\vec\theta_\tref$ is true, and $p(\vec\theta)$ is the
density of probability of $\vec\theta_\tref$ being equal to $\vec\theta$, prior
to the observation.  The likelihood is often expressed as the product of
independent probabilities of observing each datapoint $Y_{\tref, j}$
\cite{raba_organic_2017}
\begin{equation}
\label{eq:likelihood_expression}
p(\Yref \given \vec\theta) \propto \prod_{j=1}^{m}
\exp\left[
  -\frac{1}{2}\left(
      \frac{Y_{j}(\vec\theta) - Y_{\tref,j}}
           {\sigma_j}
  \right)^2
\right]
\end{equation}
where $\sigma_j$ quantifies the measurement uncertainty.

From Eqs.~\eqref{eq:MSE_expression}--\eqref{eq:likelihood_expression}, the
relation between these two points of view may be highlighted by considering the
logarithm of the posterior probability density
\begin{equation}
	\label{eq:relationship_MSE_lnprob}
	\log\big(p(\vec\theta \given \Yref)\big)
	= \frac{- m \cdot \mathrm{RMSE}(\vec\theta)^2}{2}
	+ \log\big(p(\vec\theta)\big)
	+ C
\end{equation}
where the RMSE weights are chosen as $w_{j} = 1/\sigma_j^2$, and $C$ is a
normalization constant independent of $\vec\theta$.

The prior distribution $p(\vec\theta)$ expresses the knowledge already available
before the observation of $\Yref$. Such knowledge may originate from physical
constraints or previous measurements, and generally lacks a sharp peak,
reflecting one's ignorance about the true value of $\vec\theta$.  In contrast,
the posterior distribution $p(\vec\theta \given \Yref)$ ideally features one or
several modes for parameter choices that best explain $\Yref$.  Under these
conditions, it is apparent from Eq.~\eqref{eq:relationship_MSE_lnprob} that
minimizing the RMSE can be interpreted as a maximization of the posterior
probability.  In both cases, extremization leads to the most probable value of
$\vec\theta$, given the observation $\Yref$ and inference model~$\mathcal{M}_i$.

In addition, the probabilistic Bayesian picture stresses the relevance of
computing a \emph{credible region} in which the true $\vec\theta$ lies with a
high probability, rather than a single best value.  On the practical side this
shifts the focus away from local minimization to global sampling, which can be
achieved by state-of-the-art Markov Chain Monte Carlo (MCMC) algorithms
\cite{goodman_ensemble_2010,foreman-mackey_emcee_2013}.

Because the Bayesian picture offers a more comprehensive view of the parameter
space, this work leverages MCMC sampling for parameter estimation.  This is
achieved with a custom implementation of the \emph{emcee} Python library
\cite{foreman-mackey_emcee_2013}.  A full description of its well-established
algorithm can be found in Refs.~\onlinecite{goodman_ensemble_2010,%
foreman-mackey_emcee_2013}. Its main features are outlined here to help the
discussion.  To sample the posterior distribution $p(\vec\theta \given \Yref)$,
a set of walkers iteratively explores the parameter space, in parallel.  At each
iteration, walker positions are updated by using the stretch-move rule, a
Markovian process~\cite{foreman-mackey_emcee_2013}.
During an initial \emph{burn-in phase}, the walkers discover the search-space
while being statistically pulled towards the posterior modes.  In the stationary
regime, or \emph{sampling phase}, the walker positions obtained at every new
iteration are independent samples drawn from the posterior distribution.
For all the MCMC sampling presented in this work, 64 walkers were used.
From the resulting Markov chains, one can compute a \emph{credible interval} for
each parameter $\theta_j$.  We report the 16th--84th percentile interval of the
collected samples because it coincides with the $\mu \pm \sigma$ interval when
the posterior probability density is a normal distribution with mean $\mu$ and
variance $\sigma^2$.  It contains the true value of the parameter with a
$\SI{68}{\percent}$ probability.
Further information about the Bayesian picture of parameter estimation can be
found in the dedicated literature \cite{mackay_information_nodate,%
goodman_ensemble_2010}.

\subsection{Reference datasets and uncertainties}
\label{sec:control_datasets}

In order to demonstrate that the results discussed are independent of the \GT
choice, the analysis is repeated with two different sets of parameters,
hereafter referred to as $\vec\theta_{\tref,1}$ and $\vec\theta_{\tref,2}$.
\begin{table}%
  \caption{%
    \label{tab:params_geom}%
    Geometrical parameter settings used for the simulation of the reference
    datasets, as defined in Fig.~\ref{fig:geometries}c.
  }%
  \begin{ruledtabular}%
    \begin{tabular}{lcccccc}%
      \hfill Ground \hfill & $L$ & $S$ & $W_n$ & $W_p$ & a & $N_\mathrm{per}$ \\
      \hfill truth, $\vec\theta_{\tref,i}$ \hfill & (nm) & (mm$^2$) & (nm) &
        (nm) & (nm) & \\[2pt]
      \hline \\[-0.8em]
      1 - Symmetric & 100 & 10 & 20 & 20 & 10 & 4 \\
      2 - Asymmetric & 85 & 4.5 & 10 & 10 & 0 & \textendash
    \end{tabular}%
  \end{ruledtabular}%
\end{table}%
The former is chosen such that the donor and acceptor materials are symmetric to
one another (same physical properties).  Moreover, the tortuous conduction
pathways of the input geometry are identical for the donor and acceptor domains,
as shown in Fig.~\ref{fig:geometries}b.  This ideal situation is helpful to test
the reliability of a parameter estimation procedure.  If the effective
parameters $\vec\theta_\est$ are not symmetric, their interpretation will be
misleading because it will place undue blame for poor performances on one moiety
rather than the other.  Tables~\ref{tab:params_geom}
and~\ref{tab:results_method_naive} list the parameter values corresponding to
$\vec\theta_{\tref,1}$.

$\vec\theta_{\tref,2}$ was chosen to be similar to the values found in Table~2
of Ref.~\onlinecite{neukom_opto-electronic_2018}, for which one of the evaluated
procedures $\mathcal{P}$ was initially presented.  That procedure is therefore
expected to perform well when applied to the corresponding reference dataset.
Geometric parameters are chosen to emulate the so-called checkerboard geometry,
featuring straight conduction pathways with no tortuosity.  The parameter values
used for this \GT can be found in Tables~\ref{tab:params_geom}
and~\ref{tab:params_ref_2}.  Note that it is asymmetric as it does not have
identical moieties.

Tabulated \GT values for $\Geff$ are the spatial average of the free carriers
generation rate in 2D simulations.


The uncertainty factors $\sigma_j$ for each datapoint~$j$ are an important set
of hyper-parameters.  They represent how accurately the model is expected to
reproduce the measurement.  A natural lower bound for $\sigma_j$ is
$\sigma_{\tref, j}$, the uncertainty directly imputable to the measuring
apparatus: any attempt to reproduce the measurement beyond that accuracy limit
implies fitting the noise of the instrument.  Additional disagreement between
the reference and the prediction comes from the choice of an approximate 1D
model, which cannot reproduce the finest details of the synthetic measurements.
Failure to acknowledge this model error also leads to overfitting, and slows
down MCMC convergence considerably.

In our experience, using $\sigma_j \ee \sigma_{\tref, j}$ yields artificially
small error bars on \IV curves and impedance measurements, and in practice it is
not possible to fit the prediction to the data with that level of accuracy,
suggesting that the discrepancy is dominated by model error.  Moreover, the
noise level from the Keithley~2420 SMU is so much smaller than that of the
Agilent 4294A analyzer that the dark \CV measurement would be effectively
ignored by the procedure.
For each apparatus, we therefore apply a weighting factor $\delta_a$ to
$\sigma_{\tref, j}$.  Empirically, we found that all measurements contribute
comparable terms to the RMSE by setting $\delta_a \ee 40$ for the SMU; $\delta_a
\ee 4$ for the impedance analyzer; and $\delta_a \ee 1$ for the oscilloscope.

The whole reference datasets~1 and~2, associated with $\vec\theta_{\tref,1}$ and
$\vec\theta_{\tref,2}$, are shown as black symbols in
Figs.~\ref{fig:reference_dataset_1} and~\ref{fig:reference_dataset_2},
respectively.  The uncertainties $\sigma_j \ee \delta_a{\cdot}\,\sigma_{\tref,
j}$ used for inference are displayed as error bars in all frames, except for
transient measurement frames where the $Y_j\,{\pm}\,\sigma_j$ region is shaded
in gray.

\section{Results and discussion}

\subsection{Parameter estimation from current-voltage characteristics alone}
\label{sec:fit_IV}

Considering solely the light \IV measurement taken from the reference dataset~1,
the MCMC procedure is used to search for the values of the following parameters:
the carrier mobilities ($\mu_n$, $\mu_p$), the effective free-carrier generation
rate~$\Geff$, the bimolecular recombination rate~$\krec$, the built-in
voltage~$\Vbi$, the extraction barriers ($\phicat$, $\phian$), and the parallel
resistance~$R_p$.
The device thickness~$L$, active area~$S$, series resistance~$R_S$, and relative
permittivity~$\epsr$ are assumed to be known accurately and kept equal to their
\GT value during the extraction.
The starting points of the procedure are randomly drawn within a broad region of
realistic values centered around $\vec\theta_{\tref,1}$.
The parameters $\mu_n$, $\mu_p$, $R_p$, and $\krec$ are mapped to a logarithmic
scale (log-scale) to accelerate MCMC convergence.

\input{ms_table3.tex}

\begin{figure*}
  \centering
  \includegraphics[width=\linewidth]{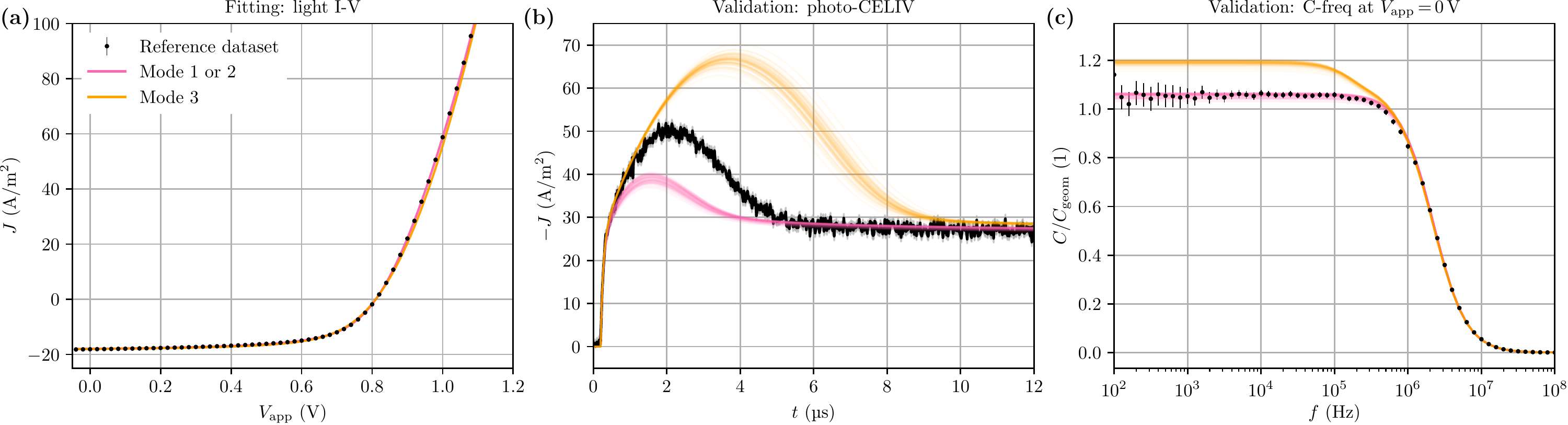}
  \caption{\label{fig:fit_IV_results}%
    Parameter extraction from light \IV curves alone may fail to generalize to
    other measurements.  Reference dataset~1 is shown with error bars (black
    symbols), along with predictions $Y(\vec\theta_\est)$ from \IV curve
    regression as a bundle of 64 curves.  a) light \IV curve (fitting); b)
    photo-CELIV transient (cross-validation); and c) \Cf response at $f \ee
    \SI{1}{\kilo\hertz}$ (cross-validation).
  }
\end{figure*}

Three local minimum of I-V curve's RMSE (i.e.~modes) are detected after
\num{1000}~iterations.  Sampling such a multimodal distribution can be slow
using \emph{emcee}.  Therefore, credible intervals were obtained by sampling
each mode separately for an additional \num{1000} iterations.  The results are
summarized in Table~\ref{tab:results_method_naive}.

As $\mu_n$ and $\mu_p$ show strong correlations, the variables
$\sqrt{\mu_n{\vphantom{l}\times}\mu_p}$ and $\sqrt{\mu_n{/}\mu_p}$ are
considered instead.  In log-scale, they relate to the original variables through
an affine transformation, hence \emph{emcee} sampling is not affected by this
change~\cite{foreman-mackey_emcee_2013}. 
Modes~1 and 2 are broad and feature imbalanced physical properties, but are
symmetric to each other.  Therefore, the parameters of mode~2 are deduced from
mode~1 by swapping the roles of the two moieties (donor and acceptor). Due to
the same intrinsic symmetry of $\mathcal{M}_i$, modes~1 and~2 give exactly the
same response $Y(\vec\theta)$. Hence, mode~2 is not discussed further.  In
contrast, mode~3 features balanced mobilities, though an order magnitude
smaller, and a recombination rate 34 times smaller than Langevin theory.

Fig.~\ref{fig:fit_IV_results}a shows that the \IV characteristics at the last
MCMC iteration for modes~1 (pink bundle of curves) and~3 (orange bundle) agree
visually with the characterization data exploited for fitting (black symbols).
In spite of this apparent success, it is clear from
Table~\ref{tab:results_method_naive} that the inferred parameters are not
reliable: estimates from mode~1 and~2 are so broad that the error bars allow for
only one significant figure on mobilities, recombination rate or parallel
resistance.
Besides, the lowest RMSE is observed for mode~1, which features imbalanced
carrier mobilities and extraction barriers.  Interpretation of the device's
performance in terms of these effective parameters is therefore misleading, as
one may incorrectly conclude that the next best experimental step is to focus on
hole transport, wasting time and resources.  It is clear in this numerical
experiment that neither moiety intrinsically performs worse than the other.
Mode~3 captures the symmetry of the \GT parameters, but its RMSE is two orders
of magnitude larger than mode~1.

To discriminate the modes, the parameters can be validated against other
characterizations of the reference.  The final walkers positions are taken as
input parameters to generate responses for each control measurement listed in
Table~\ref{tab:characterization_list}.  Since it has a very small computational
cost, this validation is advisable whenever the characterization data is
available.  Results are shown on Fig.~\ref{fig:fit_IV_results}b and
Fig.~\ref{fig:fit_IV_results}c for photo-CELIV and \Cf curves, respectively.
Clearly, none of these modes manage to appropriately fit all measurements at
once.
Another approach is therefore necessary to yield the physical parameters.

\subsection{Parameter estimation from multiple characterizations}
\label{sec:Neukom_evaluation}

A natural remedy to the concerns raised in Sec.~\ref{sec:fit_IV} would be to
work with 2D or 3D models for inference tasks to benefit from an explicit
morphology description.  Although this solution has been employed successfully
in the past, the increase in computation times is so large that it is too
time-consuming in many practical cases~\cite{raba_organic_2017}.

As an alternative, several authors have suggested using more than one
measurement to constrain the fit and avoid misleading interpretations based on
\IV curves alone~\cite{set_what_2015, neukom_opto-electronic_2018}.
Neukom~\etal~have recently illustrated the use of a broad set of measurements
for this task~\cite{neukom_opto-electronic_2018}. They outlined a seven-step
estimation procedure by associating parameters to the measurement from which
they are most easily deduced, as stated in Table~\ref{tab:neukom_steps}. The
first six steps, from A to F, provide the first estimate of $\vec\theta$, which
then serves as a starting point to the global fitting step~G.

\begin{table}[b]%
  \caption{%
    \label{tab:neukom_steps}%
    Summary of the parameter extraction steps of the procedure proposed in
    Ref.~\onlinecite{neukom_opto-electronic_2018}.
  }%
  \begin{ruledtabular}%
    \begin{tabular}{lll}%
      Step & Characterization data & Extracted parameters \\[1pt] \hline \\[-0.9em]
      \multirow{2}{*}{A} & dark-CELIV & $\epsr$ \\
           & \Cf    & $R_S$ \\
      B    & dark \IV                 & $R_p$  \\
      C    & light \IV                & $\Geff$ \\
      D    & TPC                      & $\mu_n$, $\mu_p$ \\
      E    & light \IV, dark \CV      & $\phicat$, $\Vbi$, $\phian$ \\
      F    & photo-CELIV              & $\gamma$ \\
      G    & all of the above         & $\mu_n$, $\mu_p$, $\phicat$, $\Vbi$, $
      \phian$, $\gamma$
    \end{tabular}%
  \end{ruledtabular}%
\end{table}%

\begin{figure*}
  \centering
  \includegraphics[width=\linewidth]{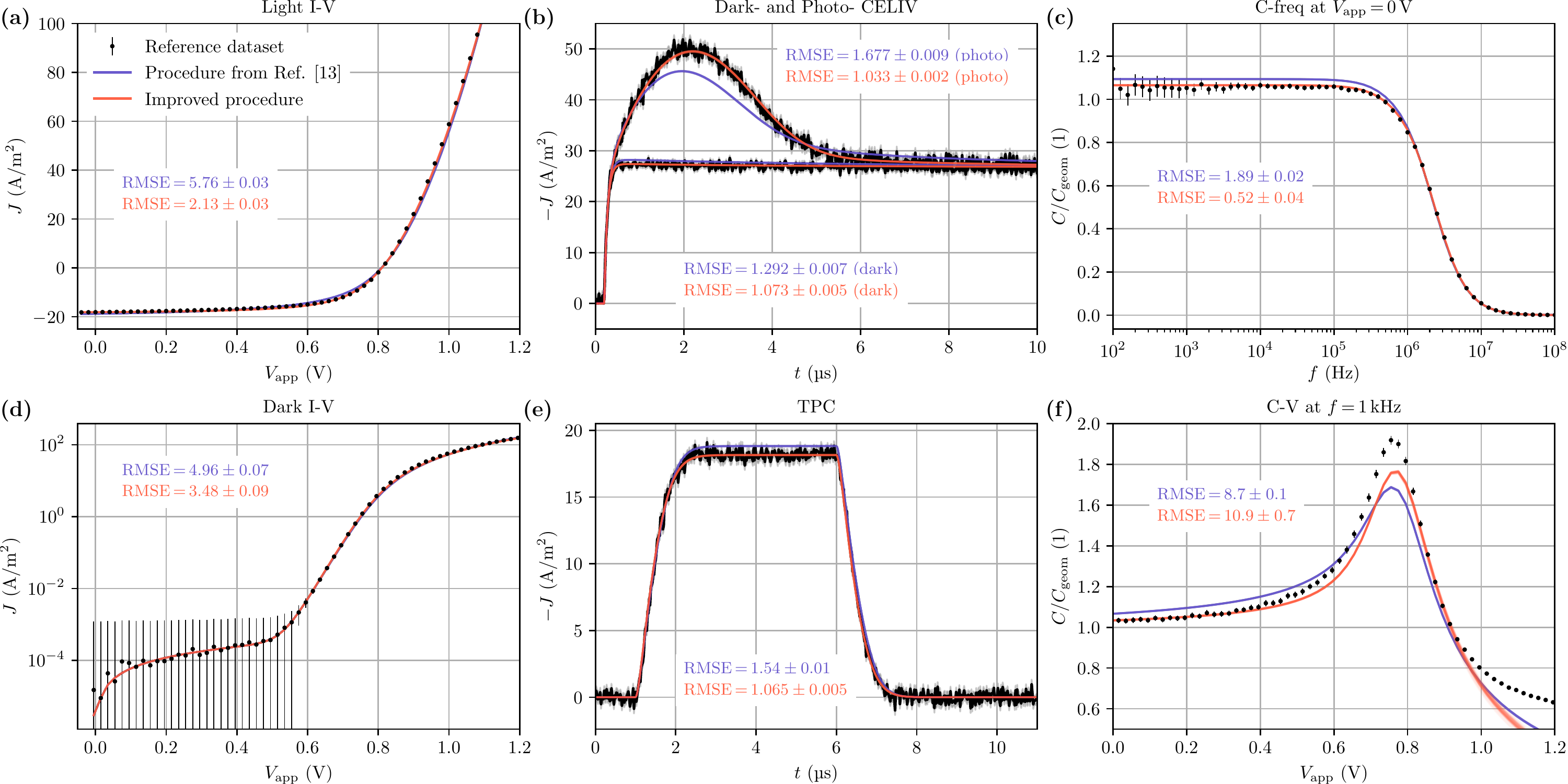}%
  \caption{%
    \label{fig:reference_dataset_1}%
    Reference dataset~1 (black symbols), along with predictions from the
    procedure taken from Ref.~\onlinecite{neukom_opto-electronic_2018} (blue
    curves) and our proposed procedure (red curves).
    Fitting results are shown as a bundle of 64 predictions
    $Y(\vec\theta_\est)$, obtained from the last MCMC iteration.  Bundle spread
    illustrates the uncertainty on $\vec\theta_\est$, propagated to
    $Y(\vec\theta_\est)$.  The uncertainties $\sigma_j \ee
    \delta_a{\cdot}\,\sigma_{\tref,j}$ used for inference are displayed as a
    gray region in frames (b) and (e) and as error bars elsewhere.
    A RMSE of $1$ indicates that the prediction error is of the order of
    $\sigma_j$.
  }%
\end{figure*}

\begin{figure*}[t]
  \centering
  \includegraphics[width=\linewidth]{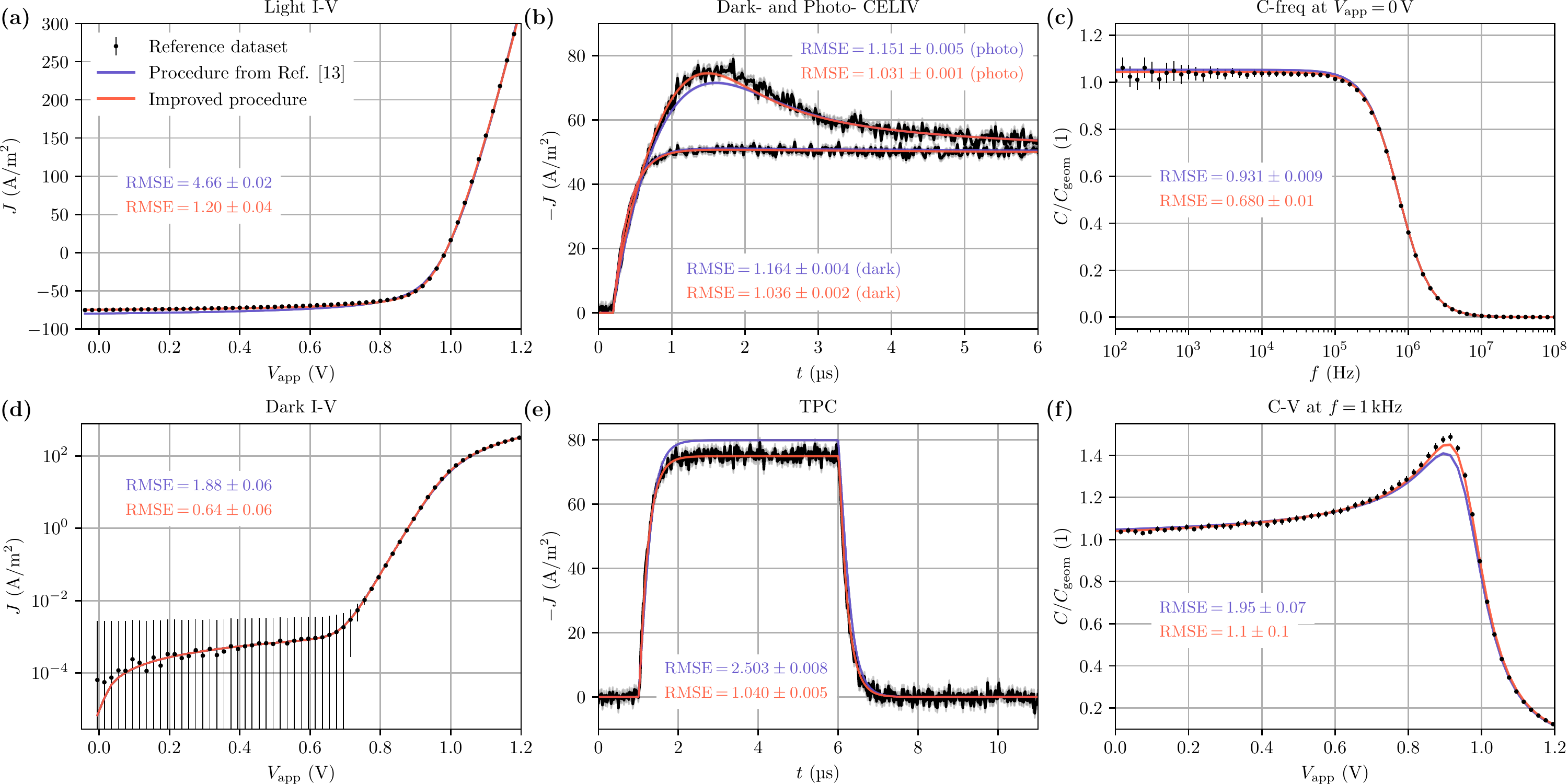}
  \caption{%
    \label{fig:reference_dataset_2}%
    Reference dataset~2 (black symbols), along with predictions from the
    procedure taken from Ref.~\onlinecite{neukom_opto-electronic_2018} (blue
    curves) and our proposed procedure (red curves).
    Fitting results are shown as a bundle of 64 predictions
    $Y(\vec\theta_\est)$, obtained from the last MCMC iteration.  Bundle spread
    illustrates the uncertainty on $\vec\theta_\est$, propagated to
    $Y(\vec\theta_\est)$.
    The uncertainties $\sigma_j \ee \delta_a{\cdot}\,\sigma_{\tref,j}$ used for
    inference are displayed as a gray region in frames (b) and (e) and as
    error bars elsewhere.
    A RMSE of $1$ indicates that the prediction error is of the order of
    $\sigma_j$.
  }
\end{figure*}

In this section, the robustness of this procedure is evaluated.  In addition, we
discuss important settings that are not explicitly reported or justified in
Ref.~\onlinecite{neukom_opto-electronic_2018}.  As mentioned in
Sec.~\ref{sec:methodo}, the evaluation protocol is applied to the two reference
datasets shown in Figs.~\ref{fig:reference_dataset_1}
and~\ref{fig:reference_dataset_2}.
Steps~A, B, C, and~F use local fitting approaches,
whereas steps~D and~E are performed over \num{200} MCMC iterations,
and the global fitting step~G is performed over \num{500} MCMC iterations.

\subsubsection{Ground-truth 1}

\paragraph{Step~A} The plateau current value $I_\infty$ of the dark-CELIV
measurement (Fig.~\ref{fig:reference_dataset_1}b) is directly related to the
geometric capacitance $\Cgeom$. If the aspect ratio $S/L$ is known, the relative
permittivity $\epsr$ can be deduced from the relation:
\begin{equation}\label{eq:I_infty}
    I_{\infty} = \mathcal{A} \cdot \Cgeom
               = \mathcal{A} \cdot \varepsilon_0 \epsr \cdot S / L
\end{equation}
where $\mathcal{A}$ is the slope of the CELIV  voltage ramp.

A slope $\mathcal{A} \ee \SI{- 100}{\volt\per\milli\second}$ was applied during
\SI{30}{\micro\second}. The current is averaged over the
\SIrange{20}{30}{\micro\second} interval, where it has reached saturation.  At a
sampling rate of \SI{100}{\points\per\micro\second}, this provides enough
statistics to cancel the noise from the apparatus.

We obtain $\Cgeom \ee \SI{2.656 \pm 0.002}{\nano\farad}$ from the data, and if
$L$ and $S$ are known perfectly, this gives directly $\epsr \ee \SI{3.000 \pm
0.002}{}$ which matches the \GT.  Otherwise, the uncertainties from $L$ and $S$
must be propagated to the error bar of $\epsr$.
Because a 1D model is used here, one should keep in mind that the value of
$\epsr$ extracted is only an effective value, averaged over the entire junction.

With $\Cgeom$ known, the capacitance of the device in the high-frequency region
(Fig.~\ref{fig:reference_dataset_1}c) is then fitted to
\begin{equation}\label{eq:cfreq_model}
  C(\omega) = \frac{\Cgeom}{ 1 + (\omega / \omega_c)^2}
\end{equation}
where the cutoff frequency $\omega_c$ depends on $R_S$ according to the
expression
\begin{equation}\label{eq:cutting_frequency}%
  \omega_c = 1 / (R_S \cdot \Cgeom).
\end{equation}
Here, a series resistance of $R_S = \SI{20.1}{\ohm}$ is found, in good agreement
with the GT value $R_S = \SI{20}{\ohm}$.

\paragraph{Step~B} The Ohmic regime of the dark \IV curve
(Fig.~\ref{fig:reference_dataset_1}d) is fitted to estimate the parallel
resistance $R_p$, yielding $R_p = \SI{167.3}{\mega\ohm}$, again in reasonable
agreement with the \GT.

\paragraph{Step~C} The effective free-carrier generation rate $\Geff$ is
adjusted to reproduce the short-circuit current $\JSC$ at one sun
(Fig.~\ref{fig:reference_dataset_1}a).
A simple bisection algorithm is efficient here because only one parameter is
extracted, and $\Geff$ is bounded by the incoming photon flux $\Gopt$.  Ten
iterations are sufficient to reach a precision of $\SI{0.1}{\percent}$ of
$\Gopt$, yielding $\Geff = \SI{1.267 \times 10^{21}}{\GUnit}$, a moderate
$\SI{6}{\percent}$ overestimation from the \GT.

\paragraph{Step~D}
\label{par:31D}

\begin{figure}[b]%
  \centering
  \includegraphics[width=0.7\linewidth]{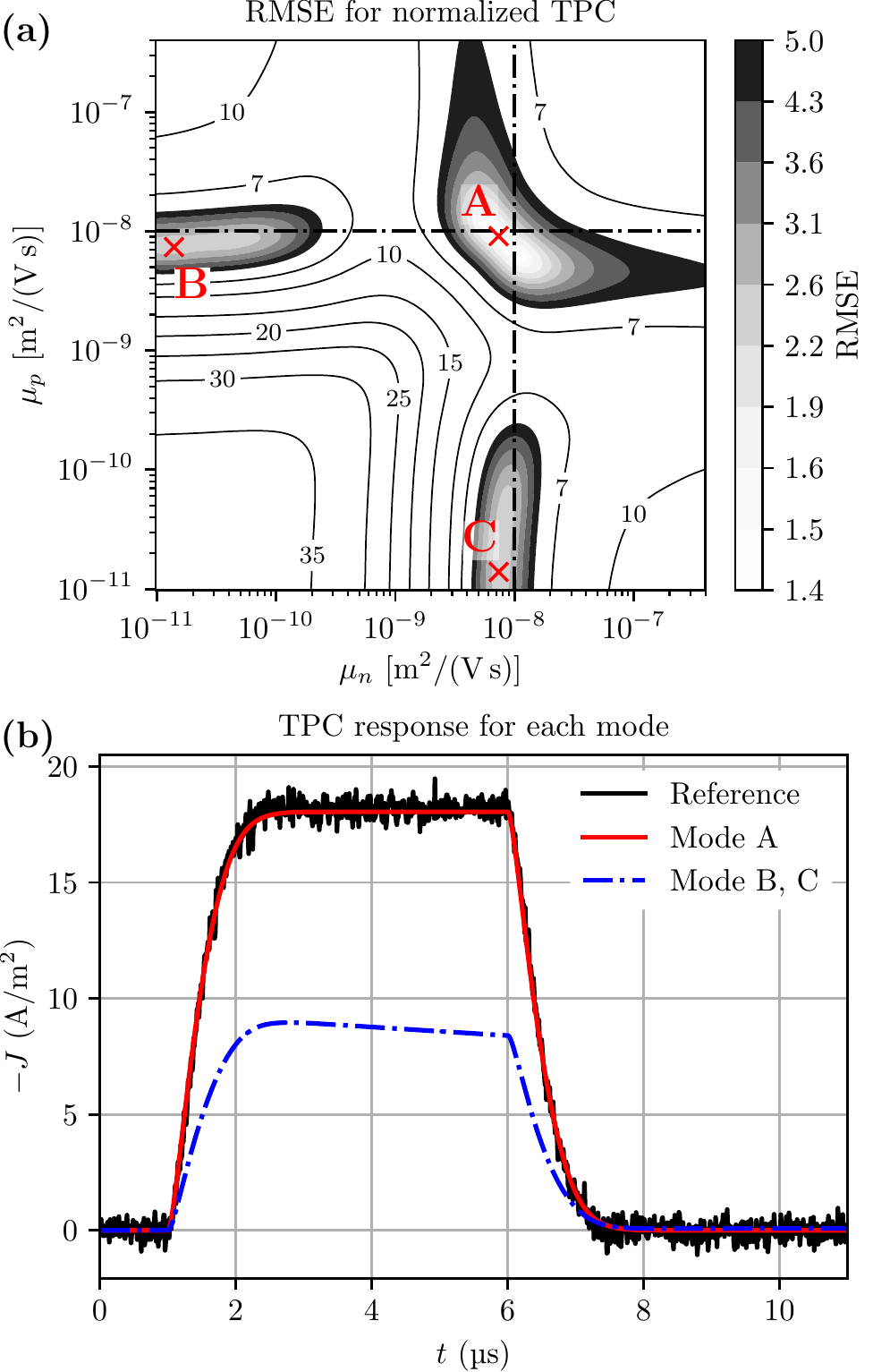}
  \caption{%
    \label{fig:mobility_3_modes}%
    a) $\mathrm{RMSE}(\vec\theta)$ landscape at step~D when TPC response is
    normalized.  Local minima are denoted by letter-labeled red crosses.  The
    black dash-dotted reticle indicates the position of the ground-truth;
    b) Corresponding TPC response for each local minimum.  Notice that minima B
    and C are superimposed because swapping moieties does not affect output
    currents.
  }%
\end{figure}%

The electron and hole mobilities are extracted from the normalized TPC
measurements (Fig.~\ref{fig:reference_dataset_1}e).  In order to construct a
complete picture of the parameter space, we sample it with our MCMC procedure.
Here again, $\mu_n$ and $\mu_p$ are set to vary in log-scale because this
produces much faster convergence in our experience.

Fig.~\ref{fig:mobility_3_modes}a shows the RMSE landscape as a function of
($\mu_n$, $\mu_p$), obtained by interpolating over a broad sampling of the
plane.  Three local minima are indicated with red crosses, while the black
reticle indicates the \GT values used to generate the reference data.  After the
Markov chains have stabilized, a dominant mode with balanced mobility is found
for $\mu_n \ee \SI{(8.2 \pm 0.3){\times}10^{-8}}{\muUnit}$ and $\mu_p \ee
\SI{(8.1 \pm 0.3){\times}10^{-8}}{\muUnit}$ (A-label in
Fig.~\ref{fig:mobility_3_modes}).  It is also the global RMSE minimum.
The extracted mobilities underestimate the \GT by about \SI{20}{\percent} but,
more importantly, the symmetry issues raised in Sec.~\ref{sec:fit_IV} no longer
correspond to the dominant modes.

Normalizing TPC implies that the dependence of $\JSC$ on carriers mobility will
not affect the fitting agreement.  In principle, this allows step~D to focus on
the time-dependence of the current rise and decay.  However it may yield
surprising results when using a local minimization procedure such as
Levenberg-Marquardt~\cite{levenberg_method_1944,marquardt_algorithm_1963,
more_levenberg_1978}.  Indeed, in addition to the dominant (balanced) mode A,
Fig.~\ref{fig:mobility_3_modes}a reveals the presence of two spurious modes (red
crosses labeled B and C) with imbalanced mobilities, separated by almost three
decades.  If a local fitting procedure is initialized in the proximity of these
two modes, the extracted mobilities may be biased.

Although the RMSE has a clear attraction sink for all three modes if the
transient currents are normalized, the current density of the TPC plateau is
underestimated by the two spurious modes by more than a factor of two, as shown
on Fig.~\ref{fig:mobility_3_modes}b.  Clearly, they can only exist if one chooses
to normalize the currents, which seems counter-productive.

\paragraph{Step~E} For this step, the built-in voltage $\Vbi$, along with
extraction barriers $\phicat$ and $\phian$, is extracted from \IV and \CV
characteristics.
In their article \cite{neukom_opto-electronic_2018}, Neukom~\etal~stressed the
importance of \IV curve's first quadrant ($V \,{>}\, 0, J \,{>}\, 0$), as it may
reveal the presence of extraction barriers.  Therefore, we ramp the voltage from
\SIrange{0}{1.5}{\volt}, roughly twice the reference's open-circuit voltage
($\VOC$), as shown on Figs.~\ref{fig:reference_dataset_1}a
and~\ref{fig:reference_dataset_1}f.

During this step, only the $\VOC$ and the \CV peak position ($\Vpeak$) are
adjusted.  We noticed that they both vary linearly with the sum $\Vgap
\,{\equiv}\, \Vbi + \phicat + \phian$ in the range of parameter values visited
by MCMC.
Therefore, the procedure uses the additional degrees of freedom (two out of
three) to attempt to recover fine details of the \IV and \CV curve's injection
regimes, which are subject to a strong model error.  As a result, the barrier
values obtained at this stage appear to be arbitrary, and it is more meaningful
to extract the value of $\Vgap$.  We obtain $\Vgap \ee \SI{1.4934 \pm
0.0001}{\volt}$, a~\SI{+7}{\percent} deviation from the \GT.  The error bar is
two orders of magnitude smaller than $k_B T/q \simeq \SI{26}{\milli\volt}$, the
smallest voltage appearing in the model.  This reveals how sensitive the RMSE is
to $\Vgap$, a consequence of the magnitude of the SMU uncertainty around $\VOC$
($\sim \SI{0.1}{\micro\ampere}$).

\paragraph{Step~F} For this step, the recombination prefactor $\gamma$ is varied
in order to match the photo-CELIV measurements
(Fig.~\ref{fig:reference_dataset_1}b).  Because there is only one degree of
freedom, it is sufficient to perform a brute-force search by varying $\gamma$
logarithmically from \numrange{10^{-2}}{10^{+2}}.  Within this range, the RMSE
exhibits a clear minimum at $\gamma \ee \num{0.293}$, about three times lower
than Langevin recombination ($\gamma \ee \num{1}$).

\paragraph{Step G} For this step, a global fitting procedure is run using MCMC
sampling.  As shown by the blue bundle of curves in
Fig.~\ref{fig:reference_dataset_1}, a relatively good agreement is obtained for
nearly all measurements involved in the fitting procedure.
However, the low-frequency capacitance or the \CV peak magnitudes are not
reproduced within the error bars.  Likewise, while the time-scales involved in
the CELIV measurements are correctly reproduced, the height of the photo-CELIV
peak is underestimated by the model.  Finally, the decrease in capacitance in
the injection regime ($V > \Vpeak$) is found to be consistently steeper in the
1D model than the 2D reference.  These discrepancies result from the 1D model
being unadapted to describe the reference.  As a result, the fitting step is
forced to accept contradictory compromises between different measurements.

Despite the uneven agreement, one can notice from Table~\ref{tab:params_ref_1}
that the estimations of the energy levels have drastically improved with respect
to step~E.  The new estimate $\Vgap \ee \SI{1.4702 \pm 0.0002}{\volt}$ is
slightly closer to the \GT (\SI{+5}{\percent}).  Finally, the mobilities have
remained balanced but are now underestimated by as much as \SI{30}{\percent}.
The full list of extracted parameters can be found in
Table~\ref{tab:params_ref_1}.

\input{ms_table5.tex}
\paragraph{Cross-validation} In Ref.~\onlinecite{neukom_opto-electronic_2018},
the inferred parameters are cross-validated by showing that the model's
predictions are in excellent agreement with TPV measurements, although not
exploited for fitting.  However, we find that a good agreement on TPV is reached
already at step D, well before the procedure's end, while the parameters are
still evolving significantly.  This suggests that the TPV response is much
easier to reproduce than other measurements, such as \CV or CELIV.  In our
experience, a set of parameters providing a satisfactory agreement at step~G
will always have a RMSE close to \num{1} on TPV, whereas the opposite is clearly
not true.  Therefore, the success of this cross-validation step should not
significantly improve the credibility of the inferred parameters.

\subsubsection{Ground-truth 2}

It will now be shown that our main conclusions are not tied to a specific choice
of \GT, as they remain valid for~$\vec\theta_{\tref,2}$ and the reference
dataset~2.

\paragraph{Steps~A to~C} Up to step~C, no additional observation is to be made.
The results from these steps are summarized in Table~\ref{tab:params_ref_2} and
confirm the precision and accuracy of the estimations obtained with the
reference dataset~1.

\paragraph{Step~D} As can be noted from Table~\ref{tab:params_ref_2}, the
average mobility $\sqrt{\mu_n{\vphantom{l}\times}\mu_p}$ of the \GT is
accurately reproduced, even though the error bars on estimates are much larger
for this reference.  The estimations of $\mu_n$ and $\mu_p$ are balanced, in
spite of the mobility ratio $\mu_n/\mu_p = 2$ associated with the \GT.  As for
the reference~1, the time-scales of the TPC are well reproduced
(Fig.~\ref{fig:reference_dataset_2}e).

\paragraph{Step~E} This step confirms that the RMSE is solely controlled by
$\Vgap$.  The Markov chains yield the value $\Vgap \ee
\SI{1.68417\pm\,0.00006}{\volt}$, which is again excessively precise, while it
overestimates the \GT by ca.~\SI{7}{\percent}. The positions of $\VOC$ and
$\Vpeak$ are correctly reproduced, but $\phicat$ and $\phian$ are estimated with
only one significant figure since this stage fails to constrain the barriers.

\paragraph{Step~F} Again, step~F is performed using a brute force search, in
which $\gamma$ varies logarithmically from \numrange{10^{-2}}{10^{+2}},
minimizing the RMSE at $\gamma \ee \num{0.215}$.

\paragraph{Step~G} Fig.~\ref{fig:reference_dataset_2} shows that the prediction
of the last iteration of the global fitting step~G (blue bundle) accurately
reproduces the broad set of reference measurements (black symbols).

Impedance measurements (see Figs.~\ref{fig:reference_dataset_2}c and
\ref{fig:reference_dataset_2}f) are better reproduced for this reference dataset
than for the previous one.  In contrast, the short-circuit current (and
therefore the TPC plateau) is overestimated, as visible on
Figs.~\ref{fig:reference_dataset_2}a and~\ref{fig:reference_dataset_2}e, while
the height of the photo-CELIV peak is underestimated.  We attribute this to the
fact that $\Geff$ was fixed early in the procedure (see
Table~\ref{tab:neukom_steps}).  Because $\JSC$ has a strong impact on the
overall RMSE, this likely hinders the procedure's progress on other parameters
as well.

The full list of extracted parameters can be found in
Table~\ref{tab:params_ref_2}.  In particular, the estimate of $\Vgap$ improves
(\SI{4}{\percent} overestimation), but the individual barriers $\phicat$ and
$\phian$ are far from the \GT.  A mobility ratio as large as $\mu_n/\mu_p = 50$
is found, which severely exaggerates the difference between electron and hole
transport.

\input{ms_table6.tex}
\subsection{Improved parameter estimation procedure}
\label{sec:improved_procedure_evaluation}

Based on the observations made so far, we now suggest changes to
Neukom~\etal{}'s procedure~\cite{neukom_opto-electronic_2018}.  The proposed
modifications seek to address major issues raised in the previous section.
Compared to the steps summarized in Table \ref{tab:neukom_steps}, the main
alteration is to discard steps~E and~F while releasing $\Geff$ at steps~C, D and
G.

\subsubsection{Step-by-step description of the procedure}

\paragraph{Steps A to C} As steps A, B and C are already providing adequate
results, they were left unchanged, resulting in the same fitting quality on
$\epsr$, $R_S$ and $R_p$.
However, keeping $\Geff$ fixed afterwards is somewhat arbitrary since $\JSC$ is
dependent on the mobilities and the recombination rate, which are only extracted
at a later stage.  Hence, $\Geff$ is adjusted in subsequent steps.  An
alternative could be to extract $\Geff$ from the saturation current at high
reverse bias, which is less sensitive to mobilities and recombination rate than
$\JSC$.

\paragraph{Step D} The main effect of $\Geff$ is to linearly rescale the entire
TPC response, but it has a little-to-no influence on the rise and decay times.
Releasing this additional parameter should thus be sufficient to reproduce the
plateau current at steady-state.  Therefore, the TPC is no longer normalized at
this stage.  Combined, we expect that these two alterations are effective at
removing the spurious modes of Fig.~\ref{fig:mobility_3_modes}.

\paragraph{Steps~E and~F}  We found that extraction barriers estimated from
step~E are not reliable, and since $\Vgap$ is easy to recover during the global
fit because it correlates linearly with $\VOC$, we suggest skipping this stage.
One could expect step~F to accelerate the convergence of the global fit as the
estimation of $\gamma$ is stable when moving from steps~F to~G.  In practice, we
found that this convergence boost was not significant, and we propose to ignore
this step also.

\paragraph{Step G} By monitoring the RMSE of each measurement from steps~C to~G,
we noticed that some characterizations have a RMSE negligible compared to the
total before they even participate in the fit.  It suggests that they only
contain information that is already captured by other measurements.  For
instance, the dark-CELIV is dominated by RC effects, also probed by the
photo-CELIV.  It makes dark-CELIV information redundant in an extraction stage
where the photo-CELIV is already in use.
Likewise, the dark \IV curve repeats information contained in the light \IV
curve, except for the estimation of $R_p$ at step~B.
Impedance measurements were not found to be redundant with any other technique,
but in our experience, it is challenging to obtain a low RMSE on both transient
and impedance measurements using the 1D model (see
Sec.~\ref{sec:control_datasets}).

As a result, we suggest using a reduced step~G', in which only light \IV, TPC,
and photo-CELIV measurements contribute to the RMSE.  We expect these
modifications to only have a minimal impact on the inference results while
allowing faster convergence.  Another approach, not explored in this work, would
be to expand the effective 1D model in ways that allow to simultaneously
reproduce impedance measurements and transient measurements equally well.

Restraining the number of fitting measurements in this way has three main advantages:
1) it reduces the fitting time by reducing the number of simulations run at each
iteration;
2) convergence is expected to be reached in fewer iterations because
contradictory requirements between transient and impedance responses are lifted;
3) the discarded measurements can either be spared (reducing the number of
experiments required) or used to expand the cross-validation dataset.

Indeed, we argue that we achieve strong cross-validation of the inferred
parameters by using $\vec\theta_\est$ to predict the dark \IV, dark-CELIV, \Cf,
dark \CV, and TPV curves.
Among these five measurements, the \Cf and dark \CV curves are the most
informative because they are not redundant with any other characterization.  A
small RMSE for these measurements is therefore indicative of a good fit, but it
may not always be possible to achieve with a 1D model.  At least, $\Vpeak$
should be predicted as accurately as $\VOC$.
In contrast, the dark \IV, dark-CELIV, and TPV curves are easier to reproduce.
A large RMSE for these measurements would therefore suggest overfitting, but a
small RMSE would be inconclusive.

During our reduced step~G', the TPC characteristic is left without
normalization, and $\Geff$ is still released as a fitting parameter.  Parameters
$\mu_n$, $\mu_p$, and $\gamma$ are again varied logarithmically because this
reduces the number of iterations required for convergence.

\subsubsection{Ground-truth 1}

The proposed procedure is applied to the reference dataset~1.  As stated above,
and shown in Table~\ref{tab:params_ref_1}, the values for $\epsr$, $R_S$, and
$R_p$ are unchanged compared to the previous procedure.

After convergence has been reached for step D, the mobilities are underestimated
by about \SI{20}{\percent}, with balanced electron/hole transport.  While $\Geff$
increases by less than \SI{1}{\percent} compared to step~C, the two spurious
modes which failed to describe the TPC plateau have disappeared.

The predictions $Y(\vec\theta)$ obtained at the last MCMC iteration for step~G'
are displayed with red bundles of curves in Fig.~\ref{fig:reference_dataset_1}.
As can be noted, the agreement with the reference~1 is satisfactory across all
measurements except the dark \CV curve (Fig.~\ref{fig:reference_dataset_1}f).

Among the common dataset used by both procedures, it can be noted that the peak
of the photo-CELIV is reproduced more accurately using our approach
(Fig.~\ref{fig:reference_dataset_1}b).  We believe that this improvement was
enabled by lifting the strong constraint of fitting transient and impedance
measurements simultaneously.  The RMSE of the light \IV
(Fig.~\ref{fig:reference_dataset_1}a), photo-CELIV
(Fig.~\ref{fig:reference_dataset_1}b), and TPC
(Fig.~\ref{fig:reference_dataset_1}e) curves are all smaller than those from
Neukom~\etal's procedure.

Even though they were not part of the fitting dataset, the dark \IV
(Fig.~\ref{fig:reference_dataset_1}d), dark-CELIV
(Fig.~\ref{fig:reference_dataset_1}b), and \Cf
(Fig.~\ref{fig:reference_dataset_1}c) characteristics are all well reproduced.
Moreover, their RMSE is reduced compared to Sec.~\ref{sec:Neukom_evaluation}. On
the other hand, the dark \CV prediction still exhibits a marked deviation in the
injection regime (Fig.~\ref{fig:reference_dataset_1}f).  This deviation is
similar to the one already described in Sec.~\ref{sec:Neukom_evaluation}, but it
is more pronounced, which leads to a larger RMSE for that measurement overall.

$\Geff$ slightly improves during step~G', with respect to step~D.  The average
mobility was underestimated by about \SI{10}{\percent}, closer to the \GT than
the method from Ref.~\onlinecite{neukom_opto-electronic_2018} (\SI{30}{\percent}),
whilst slightly unbalanced ($\mu_p/\mu_n \,{\simeq}\, 1.32$).  As can be noted
from Table~\ref{tab:params_ref_1}, the inferred values of $\phicat$ and $\phian$
are strongly asymmetric, in spite of the symmetry of the \GT.  This corroborates
the observation that $\phian$ and $\phicat$ are not independent parameters, but
that the model is mainly sensitive to the aggregate parameter $\Vgap \ee \phicat
\,{+}\, \Vbi \,{+}\, \phian$.
While we find $\phian \ee \SI{203\pm 3}{\milli\volt}$ in close agreement to the
\GT,  this appears to be coincidental, as it has no reason to be better
reproduced than $\phicat$, for which we find a \SI{100}{\milli\volt} deviation
from the \GT.  However, if the analysis is limited to $\Vgap$ for which the
extraction is the most sensitive, $\Vgap \ee \SI{1.4598 \pm 0.0001}{\volt}$,
which overestimates the true value of $\Vgap$ by only \SI{4}{\percent}.
Provided that $\phian$ and $\phicat$ are measured independently, the fitting
procedure presented here can therefore accurately extract the value of $\Vbi$.
The effective recombination rate is found to be reduced by a factor of \num{5}
compared to Langevin theory.
The full list of extracted parameters can be found in
Table~\ref{tab:params_ref_1}.

\subsubsection{Ground-truth 2}

Fig~\ref{fig:reference_dataset_2} shows, with a red bundle of curves, the
results of fitting the reference dataset~2 with our improved procedure, while
extracted parameters are gathered in Table~\ref{tab:params_ref_2}.  Again, the
parameter values obtained at steps~A to~C are the same as Neukom \etal{}'s
approach.

As already discussed, swapping moieties in the model $\mathcal M_i$ has no
effect on output currents.  Hence the posterior distribution obtained at step~D
must feature pairs of modes for the mobilities (possibly degenerated).  Indeed,
when a mode is found at $(\mu_n, \mu_p) \ee (\mu_1, \mu_2)$, then another is to
be found at $(\mu_n, \mu_p) \ee (\mu_2, \mu_1)$.  Without loss of generality, it
is enough to consider the case $\mu_n > \mu_p$, as before, and to perform
inferences on $\mu_n \ee \max(\mu_1, \mu_2)$ and $\mu_p \ee \min(\mu_1, \mu_2)$.

Here, $\mu_n$ and $\mu_p$ are correlated with $\Geff$, even if the latter only
varies within a $\pm \SI{3}{\percent}$ range.  In that context, it is not
appropriate to define consistent credible regions for $\Geff$, $\mu_n$, and
$\mu_p$ by the direct read of the percentile intervals.
It is important to realize that due to correlations, independent values of
$\Geff$, $\mu_n$ and $\mu_p$ cannot be extracted.  At this stage (step~D), this
is not an issue because the final MCMC walkers' positions can be used as the
starting positions of the next stage.  Indeed, contrarily to a local
minimization procedure, MCMC approaches are able to represent the parameter
joint distributions in their complexity.  The parameter correlation will be
easier to reduce at step~G', when a larger set of measurements is considered.

Nevertheless, it may be valuable to get an approximate sense of the location
and spread of each parameter distribution.  Therefore, we first consider a
credible region $\mathcal{C}$ for $\Geff$ (16th--84th percentile interval), and
then characterize the distributions of $\mu_n$ and $\mu_p$, conditional on
$\Geff\,{\in}\,\mathcal{C}$ by fitting them to normal distributions.  Using this
scheme, we obtain values for $\Geff$, $\mu_n$ and $\mu_p$ in close agreement
with the \GT, as reported in Table~\ref{tab:params_ref_1}.  In particular we
find a mobility ratio $(\mu_n / \mu_p) \ee \num{2.1}$, in good agreement with
the value of \num{2.0} of the \GT.  Of course, the error bars on mobilities are
here conditional on the value of $\Geff$ and thus underestimate the true
uncertainties.

During step~G', the mobility estimations lose some of their accuracy and the
mobility ratio increases to \num{4}.  Because this ratio is large enough for the
two mobility modes to be clearly separated, no special treatment is needed to
analyze step~G'.  Therefore, all the procedures presented in this paper
exaggerate mobility imbalance.  The improved procedure does not solve this
issue, but it reduces its magnitude by a decade compared to
Sec.~\ref{sec:Neukom_evaluation}.

The accuracy on the estimation of $\Vgap$ is equivalent to Neukom~\etal{}'s
procedure (\SI{+4}{\percent}).  The full list of extracted parameters can be
found in Table~\ref{tab:params_ref_2}.

\subsection{Interpretation of the effective parameters}
\label{sec:interpretation}


Previous sections have shown that considering distinct characterizations for
fitting allows reaching a good agreement for most measurements.
Nevertheless, it is apparent from Tables~\ref{tab:params_ref_1} and
\ref{tab:params_ref_2} that the values of the ground-truth~$\vec\theta_{\tref}$
are  generally not contained within MCMC error bars of the infered parameters~%
$\vec\theta_{\est}$.  This demonstrates that the extracted parameters are only
effective, and should be interpreted as such.

In particular, the effective generation rate $\Geff$ is well reproduced by all
procedures considered in this work, suggesting it can straightforwardly be
interpreted as a volume averaged generation rate.

In contrast, the Langevin recombination rate prefactor $\gamma$ is consistently
observed to  be lower than \num{1}. As already pointed out by previous
theoretical work, charge carriers are protected from recombination in the 2D
model, because donor and acceptor domains are segregated whereas in the 1D
effective medium model, bimolecular recombination occurs in the bulk of the
active layer~\cite{albes_investigation_2016}.  It is then expected that the
effective $\gamma$ of the 1D model must be lower than the \GT to obtain the same
overall recombination currents.  Hence, caution is warranted when interpreting
values of $\gamma$ obtained from parameter estimations using 1D effective medium
models.


Likewise, it has been proposed~\cite{heiber_impact_2017,%
albes_investigation_2016} that transit times of free cariers are affected by the
details of the BHJ geometry.  Apparent carrier mobilities are expected to be
decreasing functions of the tortuosity.  This could explain why the apparent
mobilities obtained by fitting the TPC responses are \SI{20}{\percent} below the
\GT in the tortuous case~($\vec\theta_{\tref,1}$), but closer to the \GT in the
non-tortuous case~($\vec\theta_{\tref,2}$), as shown respectively in
Tables~\ref{tab:params_ref_1} and~\ref{tab:params_ref_2}.  A~clear rationale to
predict the effective mobilities from the geometry is missing and would require
further research.  While the effective parameters yielded by our procedure are
not closer to the GT than Ref.~\onlinecite{neukom_opto-electronic_2018}, the
global fitting step G' predicts the characterization datasets with higher
accuracy.


Values of $\Vgap$ reported in Tables~\ref{tab:params_ref_1}
and~\ref{tab:params_ref_2} reveal that $\Vgap$ is overestimated by the
procedures, even though $\VOC$ and $\Vpeak$ are well reproduced and the error
bar on $\Vgap$ is smaller than the precision needed.
This suggests that $\Vgap$ should also be considered as effective, which is
consistent with the fact that the $\VOC$ of BHJ devices is dependent on the
morphology \cite{ray_can_2012}.

Determining the meaning of these effective parameters is beyond the scope of
this contribution, but crucial in order to avoid misleading conclusions from
parameter extraction results.

\section{Summary and conclusion}
\label{sec:conclusion}

In this work, we have tested the robustness of three different fitting
strategies for OPV characterization data.  By generating synthetic but realistic
device responses, we can express clear conclusions about each method's accuracy
in a context that closely matches experimental conditions.  As fitting procedures
are ultimately about extracting parameters, this level of validation could not
have been achieved using experimental measurements, for which ground-truth
parameters are unknown.

We reiterate the literature's consensus that \IV data alone is insufficient to
draw meaningful inferences about physical parameters, as defined in standard
drift-diffusion models.  In fact, following that approach is prone to misleading
conclusions, wasting time and resources, as clearly demonstrated in
Sec.~\ref{sec:fit_IV}.

Moreover, after having carefully evaluated the procedure proposed in
Ref.~\onlinecite{neukom_opto-electronic_2018} (see
Sec.~\ref{sec:Neukom_evaluation}), we validated a modified procedure that
leverages substantial improvements (see
Sec.~\ref{sec:improved_procedure_evaluation}) to achieve a better accuracy with
fewer measurements.
In addition, using a global MCMC fitting procedure provides a better
appreciation of the validity of the results, in the form of error bars, and the
ability to detect several RMSE modes when they exist.
While a MCMC fit typically requires more model evaluations than Levenberg-Marquardt, it
can be run in parallel, on up to \num{32} processors in our case, and does not require
evaluating the model's Jacobian.  That makes our procedure suitable for high
fidelity parameter extraction at the expense of higher computational cost.
In this regard, replacing the MCMC sampling with a less model-intensive Bayesian
sampler may enable lower computation times in the future without changing the
overall structure of the procedure.

Regarding the extracted parameters obtained by the above-mentioned procedures,
they are all associated to a 1D effective medium model as it needs less
computational ressources. While the obtained values did not match the
ground-truth, these  
1D effective parameters can adequately represent a broad set of device
measurements, especially in the non-tortuous case.  Nonetheless, their values
must be carefully interpreted as aggregated information from BHJ morphology and
other physical properties.

Because a 1D model has fewer degrees of freedom than a 2D or 3D model, one can
not hope to fully disentangle the exact physical parameters and the BHJ
morphology from each other.  The missing link, i.e., a detailed knowledge of the
relationship between the morphology and the effective parameters, would be a
significant step towards the understanding of structure-to-performance
relationships in OPV cells.

  \begin{acknowledgments}
      A.P. received financial support from the MSII French doctoral school
      (ED-269). The authors wish to thank Dr.~M.~Fouesnau and Dr.~Ch.~Heinrich
      for fruitful discussions regarding Bayesian inference,
      Dr.~D.~Foreman-Mackey, Dr.~J.~Goodman and Dr. J.~Weare for helpful
      discussion regarding the MCMC implementation applied in this work.  Many
      thanks go to Dr.~M.~Neukom for his helpful feedback about our attempts to
      reproduce simulations from Ref.~\onlinecite{neukom_opto-electronic_2018}.
  \end{acknowledgments}

\input{ms_biblio_short.tex}
\end{document}

%% file: ms_table3.tex
\begin{table}[b]%
	\caption{%
    \label{tab:results_method_naive}%
    Ground-truth and inferred parameters for reference dataset~1 solely using
    light \IV curve for the fit.
    Error bars denote the credible interval obtained by MCMC sampling.
  }
  \begin{ruledtabular}%
    \begin{tabular}{lcccc}%
      \multirow{2}{*}{Parameter}
      & \multirow{2}{*}{Unit}
      & Ground
      & \multirow{2}{*}{Mode~1\,\footnote{Modes 1 and 2 are symmetric: mode~2 is
        deduced from mode~1 by exchanging donor and acceptor roles.}}
      & \multirow{2}{*}{Mode~3} \\
      
      \None
      & \None
      & truth 
      & \None
      & \None \\
      \hline \\[-0.8em]

	    $R_p$
	    & \si{\mega\ohm}
	    & $160.0$
	    & $17_{-16}^{+673}$
	    & $226_{-198}^{+1482}$ \\[1.5pt]

	    $\Geff \times 10^{-21}$
	    & \si{\GUnit}
	    & $1.199$
	    & $1.184_{-0.011}^{+0.007}$
	    & $1.172_{-0.005}^{+0.004}$ \\[1.5pt]

	    $\sqrt{\mu_n {\vphantom{l}\times} \mu_p} \times 10^8$
	    & \si{\muUnit}
	    & $1.0$
	    & $3.5_{-0.9}^{+3}$
	    & $0.27\,\pm\,0.01$ \\[1.5pt]

	    $\sqrt{\mu_n / \mu_p}$
	    & \si{1}
	    & $1.0$
	    & $2.0_{-0.4}^{+1.7}$
	    & $1.00_{-0.04}^{+0.03}$ \\[1.5pt]

	    $\phicat$
	    & \si{\milli\volt}
	    & $200$
	    & $365_{-16}^{+30}$
	    & $60_{-41}^{+53}$ \\[1.5pt]

	    $\phian$
	    & \si{\milli\volt}
	    & $200$
	    & $101_{-66}^{+71}$
	    & $70_{-49}^{+56}$ \\[1.5pt]

	    $\Vbi$
	    & \si{\volt}
	    & $1.0$
	    & $1.07\pm 0.07$
	    & $1.28_{-0.07}^{+0.06}$ \\[1.5pt]

	    $\krec \times 10^{16}$
	    & \si{\kUnit}
	    & $1.206$
	    & $3.9_{-1.8}^{+10.2}$
	    & $0.029 \pm 0.003$ \\[1.5pt]

	    \hline \\[-0.8em]

	    RMSE
	    & \num{1}
	    & \None{}
	    & $19.2^{+1.0}_{-2.7}$
	    & $1064.1^{+0.9}_{-2.1}$ \\

    \end{tabular}%
  \end{ruledtabular}%
\end{table}%

%% file: ms_table5.tex
\begin{table*}
  \caption{%
    \label{tab:params_ref_1}%
    Ground-truth~1 and inferred parameters from the reference dataset~1.
    Parameter kept fixed to their partial fitting estimation are indicated by
    dashes. The error bars correspond to the 16th--84th percentile interval of
    the Markov chains.
    When $p(\vec\theta \given \Yref)$ [Eq.~\eqref{eq:Bayes_theorem}] is the
    normal distribution,
    $\theta~\sim~\mathcal{N}(\mu,~\sigma^2)$, $\theta = \mu_{-m}^{+ p}$
    means that $[\mu-m:\mu+p]$ is the $\mu\pm\,\sigma$ interval.
  }
  \begin{ruledtabular}
    \begin{tabular}{llc*{2}{r@{\hskip-0.6em}l@{}c@{}r@{\hskip-0.5em}l}}

      \multirow{2}{*}{Parameter} & \multirow{2}{*}{Unit} & Ground
        & \multicolumn{5}{c}{$\vec\theta_{\est}$, method from Ref.~\onlinecite{neukom_opto-electronic_2018}}
        & \multicolumn{5}{c}{$\vec\theta_{\est}$, improved procedure (this work)} \\[1.5pt]
      \cline{4-8}\cline{9-13} \\[-0.8em]

      \None & \None & truth, $\vec\theta_{\tref,1}$
        & \multicolumn{2}{c}{Partial fit} & (step)
        & \multicolumn{2}{c}{Global fit}
        & \multicolumn{2}{c}{Partial fit} & (step)
        & \multicolumn{2}{c}{Global fit} \\[1.5pt]
      \hline \\[-0.8em]

      $\epsr$ & \None & $3$
        & $3.000$ & $\pm\,0.002$ &  (A)
        & \textendash &
        & $3.000$ & $\pm\,0.002$ &  (A)
        & \textendash &  \\

      $R_S$ & \si{\ohm} & $20$
        & $20.1$ & \None{} &  (A)
        & \textendash & \None
        & $20.1$ & \None{} &  (A)
        & \textendash & \None \\[1.5pt]

      $R_p$ & \si{\mega\ohm} & $160$
        & $167.3$ & \None{} &  (B)
        & \textendash & \None
        & $167.3$ & \None{} &  (B)
        & \textendash & \None \\[1.5pt]

      $\Geff \times 10^{-21}$ & \si{\GUnit} & $1.199$
        & $1.267$ & \None{} &  (C)
        & \textendash & \None
        & $1.275$ & $\pm\,0.001$ & (D)
        & $1.160$ & $_{-0.002}^{+0.001}$ \\[1.5pt]

      $\mu_n \times 10^{8}$ & \si{\muUnit} & $1.0$
        & $0.82$ & $\pm\,0.03$ &  (D)
        & $0.71$ & $\pm\,0.01$
        & $0.80$ & $\pm\,0.02$ &  (D)
        & $0.79$ & $\pm\,0.01$ \\[1.5pt]

      $\mu_p \times 10^{8}$ & \si{\muUnit} & $1.0$
        & $0.81$ & $\pm\,0.03$ &  (D)
        & $0.71$ & $\pm\,0.01$
        & $0.79$ & $\pm\,0.02$ &  (D)
        & $1.04$ & $\pm\,0.01$ \\[1.5pt]

      $\phicat$ & \si{\milli\volt} & $200$
        & $127$ & $_{-51}^{+12}$ &  (E)
        & $174$ & $_{-2}^{+1}$
        & \None & \None & \None
        & $302$ & $\pm\,1$ \\[1.5pt]

      $\phian$ & \si{\milli\volt} & $200$
        & $79$ & $\pm\,50$ &  (E)
        & $174$ & $\pm\,2$
        & \None & \None & \None
        & $203$ & $\pm\,3$ \\[1.5pt]

      $\Vbi$ & \si{\volt} & $1.0$
        & $1.30$ & $_{-0.02}^{+0.03}$ &  (E)
        & $1.123$ & $_{-0.001}^{+0.002}$
        & \None & \None & \None
        & $0.956$ & $\pm\,0.003$ \\[1.5pt]

      $\Vgap$ & \si{\volt} & $1.4$
        & $1.4934$ & $\pm\,0.0001$ &  (E)
        & $1.4702$ & $\pm\,0.0002$
        & \None & \None & \None
        & $1.4598$ & $\pm\,0.0001$ \\[1.5pt]

      $\gamma$ & \None & $1.0$
        & $0.293$ & \None{} &  (F)
        & $0.466$ & $_{-0.003}^{+0.002}$
        & \None & \None & \None
        & $0.213$ & $\pm\,0.002$ \\
    \end{tabular}
  \end{ruledtabular}
\end{table*}

%% file: ms_table6.tex
\begin{table*}
	\caption{%
	  \label{tab:params_ref_2}%
    Ground-truth~2 and inferred parameters from the reference dataset~2.
    Parameter kept fixed to their partial fitting estimation are indicated by
    dashes. The error bars correspond to the 16th--84th percentile interval of
    the Markov chains.
    When $p(\vec\theta \given \Yref)$ [Eq.~\eqref{eq:Bayes_theorem}] is the
    normal distribution,
    $\theta~\sim~\mathcal{N}(\mu,~\sigma^2)$, $\theta = \mu_{-m}^{+ p}$
    means that $[\mu-m:\mu+p]$ is the $\mu\pm\,\sigma$ interval.
  }
  \begin{ruledtabular}
    \begin{tabular}{llc*{2}{r@{\hskip-0.6em}l@{}c@{}r@{\hskip-0.5em}l}}
      \multirow{2}{*}{Parameter} & \multirow{2}{*}{Unit} & Ground
        & \multicolumn{5}{c}{$\vec\theta_{\est}$, method from Ref.~\onlinecite{neukom_opto-electronic_2018}}
        & \multicolumn{5}{c}{$\vec\theta_{\est}$, improved procedure (this work)} \\[1.5pt]
      \cline{4-8}\cline{9-13} \\[-0.8em]

      \None & \None & truth, $\vec\theta_{\tref,2}$
        & \multicolumn{2}{c}{Partial fit} & (step)
        & \multicolumn{2}{c}{Global fit}
        & \multicolumn{2}{c}{Partial fit} & (step)
        & \multicolumn{2}{c}{Global fit} \\[1.5pt]
      \hline \\[-0.8em]

      $\epsr$ \None & & $4.7$
        & $4.711$ & $\pm\,0.002$ &  (A)
        & \textendash & \None
        & $4.711$ & $\pm\,0.002$ &  (A)
        & \textendash & \None \\[1.5pt]

      $R_S$ & \si{\ohm} & $90$
        & $88.5$ & \None{} &  (A)
        & \textendash & \None
        & $88.5$ & \None{} &  (A)
        & \textendash & \None \\[1.5pt]

      $R_p$ & \si{\mega\ohm} & $160$
        & $163.1$ & \None{} &  (B)
        & \textendash & \None
        & $163.1$ & \None{} &  (B)
        & \textendash & \None \\[1.5pt]

      $\Geff \times 10^{-21}$ & \si{\GUnit} & $5.922$
        & $6.230$ & \None{} &  (C)
        & \textendash & \None
        & $6.11$ & $_{-0.02}^{+0.05}$ &  (D)
        & $5.614$ & $\pm\,0.006$ \\[1.5pt]

      $\mu_n \times 10^{7}$ & \si{\muUnit} & $1.6$
        & $1.2$ & $_{-0.4}^{+1}$ &  (D)
        & $8.2$ & $_{-0.5}^{+0.7}$
        & $1.8$ & $\pm\,0.6$ &  (D)
        & $1.87$ & $_{-0.10}^{+0.09}$ \\[1.5pt]

      $\mu_p \times 10^{7}$ & \si{\muUnit} & $0.8$
        & $1.2$ & $_{-0.4}^{+0.9}$ &  (D)
        & $0.163$  & $\pm\,0.002$
        & $0.9$ & $\pm\,0.1$ &  (D)
        & $0.47$ & $\pm\,0.02$ \\[1.5pt]

      $\phicat$ & \si{\milli\volt} & $80$
        & $84$ & $_{-47}^{+28}$ &  (E)
        & $46$ & $_{-3}^{+4}$
        & \None & \None & \None
        & $170$ & $\pm\,3$ \\[1.5pt]

      $\phian$ & \si{\milli\volt} & $150$
        & $81$ & $_{-49}^{+30}$ &  (E)
        & $214$ & $\pm\,1$
        & \None & \None & \None
        & $176$ & $\pm\,3$ \\[1.5pt]

      $\Vbi$ & \si{\volt} & $1.34$
        & $1.53$ & $_{-0.02}^{+0.02}$ &  (E)
        & $1.380$ & $_{-0.004}^{+0.003}$
        & \None & \None & \None
        & $1.292$ & $_{-0.003}^{+0.002}$ \\[1.5pt]

      $\Vgap$ & \si{\volt} & $1.57$
        & $1.68417$ & $\pm\,0.00006$ &  (E)
        & $1.6400$ &  $_{-0.0002}^{+0.0005}$
        & \None & \None & \None
        & $1.6371$ & $_{-0.0004}^{+0.0003}$ \\[1.5pt]

      $\gamma$ & \None & $1.0$
        & $0.215$ & \None{} &  (F)
        & $0.051$ & $_{-0.004}^{+0.003}$
        & \None & \None & \None
        & $0.147$ & $\pm\,0.005$ \\

    \end{tabular}
  \end{ruledtabular}
\end{table*}